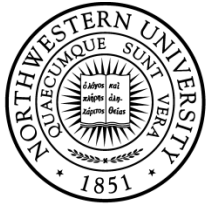

# NORTHWESTERN
## UNIVERSITY

Electrical Engineering & Computer Science Department

Technical Report
NWU-EECS-11-02
February 2, 2011

# Elastic Fidelity: Trading-off Computational Accuracy for Energy Reduction

Sourya Roy[†], Tyler Clemons[‡], S M Faisal[‡], Ke Liu[†], Nikos Hardavellas[†], Srinivasan Parthasarathy[‡]

[†]Department of Electrical Engineering & Computer Science, Northwestern University

[‡]Department of Computer Science & Engineering, Ohio State University

## ABSTRACT

Power dissipation and energy consumption have become one of the most important problems in the design of processors today. This is especially true in power-constrained environments, such as embedded and mobile computing. While lowering the operational voltage can reduce power consumption, there are limits imposed at design time, beyond which hardware components experience faulty operation. Moreover, the decrease in feature size has led to higher susceptibility to process variations, leading to reliability issues and lowering yield. However, not all computations and all data in a workload need to maintain 100% fidelity. In this paper, we explore the idea of employing functional or storage units that let go the conservative guardbands imposed on the design to guarantee reliable execution. Rather, these units exhibit Elastic Fidelity, by judiciously lowering the voltage to trade-off reliable execution for power consumption based on the error guarantees required by the executing code. By estimating the accuracy required by each computational segment of a workload, and steering each computation to different functional and storage units, Elastic Fidelity Computing obtains power and energy savings while reaching the reliability targets required by each computational segment. Our preliminary results indicate that even with conservative estimates, Elastic Fidelity can reduce the power and energy consumption of a processor by 11-13% when executing applications involving human perception that are typically included in modern mobile platforms, such as audio, image, and video decoding.

***Keywords*** Energy efficiency, hardware reliability, error tolerance, elastic fidelity

*First Workshop on Architecture and Systems Support for Mobile Applications, co-located with ASPLOS 2011*

# Elastic Fidelity: Trading-off Computational Accuracy for Energy Reduction

Sourya Roy[†], Tyler Clemons[‡], S M Faisal[‡], Ke Liu[†], Nikos Hardavellas[†], Srinivasan Parthasarathy[‡]

[†]Department of Electrical Engineering & Computer Science
Northwestern University
{souryaroy, nikos}@northwestern.edu, KeLiu2015@u.northwestern.edu

[‡]Department of Computer Science & Engineering
Ohio State University
{clemonst, faisal, srini}@cse.ohio-state.edu

**Abstract**

Power dissipation and energy consumption have become one of the most important problems in the design of processors today. This is especially true in power-constrained environments, such as embedded and mobile computing. While lowering the operational voltage can reduce power consumption, there are limits imposed at design time, beyond which hardware components experience faulty operation. Moreover, the decrease in feature size has led to higher susceptibility to process variations, leading to reliability issues and lowering yield. However, not all computations and all data in a workload need to maintain 100% fidelity. In this paper, we explore the idea of employing functional or storage units that let go the conservative guardbands imposed on the design to guarantee reliable execution. Rather, these units exhibit Elastic Fidelity, by judiciously lowering the voltage to trade-off reliable execution for power consumption based on the error guarantees required by the executing code. By estimating the accuracy required by each computational segment of a workload, and steering each computation to different functional and storage units, Elastic Fidelity Computing obtains power and energy savings while reaching the reliability targets required by each computational segment. Our preliminary results indicate that even with conservative estimates, Elastic Fidelity can reduce the power and energy consumption of a processor by 11-13% when executing applications involving human perception that are typically included in modern mobile platforms, such as audio, image, and video decoding.

***Keywords*** Energy efficiency, hardware reliability, error tolerance, elastic fidelity

## 1. Introduction

Continued technology scaling in IC design has made power dissipation a major constraint in the design of processors today. Although feature sizes are still scaling, voltage scaling has nearly stopped due to high leakage currents associated with low threshold voltages. This has lead to a dramatic increase in power density with decreasing feature size [15]. On the other hand, the scaling of the feature sizes has made chips more susceptible to problems of variability and hardware faults. These faults originate from process variations, soft errors and wear outs, hampering reliable execution [1, 17].

Traditionally, the operating points of processors have been determined by conservative guardbands based on worst-case scenarios. A guarband refers to the timing differential inserted into the hardware design to allow for signals to communicate without being perfectly aligned. However, this design approach results in significant overheads in both power and performance [22]. This leads to an interesting question: What if we let go of these guardbands and allow components of the processor to fail sometimes with the errors accommodated at the architectural and software levels? By following laws of transistor physics, keeping all else constant, decreasing the operating voltage ($V_{dd}$) would reduce power consumption at a quadratic rate, at the expense of some timing errors.

Prior research has shown that in every large CMOS chip, there exist two voltage operating points – the rated voltage point and the critical voltage point [19, 26]. This leads to three operating regions for the processor. First, when the supply voltage is at or above the rated voltage, the processor runs at full accuracy without any errors. Second, when the processor operates at a supply voltage between the rated and critical voltage points, small-scale errors emerge due to timing violations in worst-case situations. And last, operating at a voltage beyond this critical point leads to massive errors.

In this paper, we propose the idea of operating processor components (e.g., functional units) at the region of supply voltage between the rated and critical operating points, to attain significant reductions in power while meeting the reliability requirements requested by each section of the executing application. The errors originating due to this are accommodated at the software layer by exploiting the fact that different sections of the code require variable reliability guarantees to present acceptable results to the user. We envision that programming language constructs can denote the reliability guarantees required by different sections of the code; these requirements are communicated to the hardware during execution, which steers the computation to corresponding functional and storage units operating at the lowest voltage that meets the required reliability constraints. By not treating all code and all data the same from the viewpoint of reliability requirements, *Elastic Fidelity Computing* exploits sections of the computation that are error-tolerant to lower power and energy consumption, without negatively impacting executions that require full reliability.

To explore the feasibility of this idea, we examine the error tolerance of a range of applications involving human perception in the realms of audio, image and video decompression in the presence of computation errors in the ALUs. We demonstrate that:

1. Different portions of an application's dataset exhibit variable error tolerance. For example, errors occurring in the low-significance bits have a lesser effect in the application behavior than those occurring in higher-significance bits.

2. Different portions of an application's code exhibit variable error tolerance. There are some functions that show negligible effects, while others result in a program crash even if the least significant bit is flipped.

3. Our preliminary results on stock kernels (without any modifications) indicate that Elastic Fidelity Computing reduces the processor power and energy by 11-13% for our applications, even if we allow only the ALUs to exhibit Elastic Fidelity. We anticipate that expanding this idea to more execution and storage components of a processor would result in much higher power savings.

The remainder of the paper is organized as follows. Section II discusses the underlying idea of error tolerance and Elastic Fidelity Computing. Section III describes our experimental methodology while Section IV reports the results and analyses of our study. Finally, Section V presents related work and section VI concludes the paper.

## 2. Elastic Fidelity and Software Behavior

Traditionally, program execution is said to be correct if and only if the underlying computations are perfect. However, a program may still appear to execute correctly if it returns acceptable results from the user's perspective, even if there is some noise in the data or inaccuracies in the computation [23].

Prior work such as that in [23] shows that the level of error tolerance is application-dependent and depends on how accurate a program's output needs to be. There are applications which are highly resilient inherently and there are others which are very little. Important examples of highly-resilient applications come from the class of soft computing. Unlike hard or exact computing, soft computing takes advantage of the tolerance of imprecision, uncertainty and approximation for a given problem – resulting in acceptable rather than exact results [36, 37]. Multimedia applications offer a very interesting example of soft computing. These applications primarily depend on human perception and allow considerable leeway in terms of accuracy. Moreover, such applications are typically included in modern mobile platforms and are heavily exercised by users. Similarly, there are applications that already assume unreliable substrates and already have error-correcting capabilities (e.g., networking applications [24]). Other examples include Artificial Intelligence applications on forecasting, inference and data mining, scientific computing (e.g., simulations of oceanic currents, weather forecasting), or computations on already noisy data (e.g., sensor readings). Such workloads tend to perform computations on approximations, and through multiple iterations narrow down to a set of results that are within a qualitative threshold according to the user requirements. On the other hand, other applications rely on exact numerical results and are generally intolerant of errors. Examples include memory management in the operating system, code compilation and lossless data compression.

Looking further into error-resilient applications, we envisage that different portions of the execution offer different error tolerance. For example, pointer operations and control logic such as conditional and branch statements are highly sensitive to errors. A corrupt pointer would usually lead to a segmentation fault while a corrupt control would notably disorder the program execution. On the other hand, operations involving standard arithmetic computations such as matrix processing and decoding are generally error-tolerant and have a relatively benign effect on the final result of the program.

In view of these observations, we propose the idea of exploiting the elastic fidelity of computations by varying the reliability of the underlying hardware according to the application needs at each point in time. Portions of the application that are error-sensitive are executed at full reliability, while the ones that are error-tolerant are run on variable accuracy to produce an acceptable result to the end user. As discussed in the previous section, the reliability of the underlying hardware can be varied by running it on a region between the rated and critical voltage points, resulting in power savings. This scheme can be implemented at the granularity of a core, in both single and multi-core systems. In the former case, the operation of the core can be dynamically changed between full (100%) and variable accuracy, according to the requirements of the executing code segment. In the latter case, certain cores run on full accuracy to accommodate the error-sensitive operations, while others run on variable accuracy to compute error-tolerant operations. Techniques like staged execution [13] and computation spreading [4] can facilitate execution migration between the various cores. Similarly, Elastic Fidelity can be implemented at a finer granularity by varying the fidelity guarantees of individual functional units or storage elements within a single core. Fig 1 illustrates such an example with ALUs.

From the viewpoint of software design, Elastic Fidelity can be implemented through programming constructs. A programmer specifies which variables and code segments are allowable to hardware errors and their tolerance margins. In turn, the compiler maps these constructs to specialized instructions that direct the core to steer the computation to a functional unit with a specific reliability level, by changing its operating voltage. On the hardware end, dynamic voltage scaling and calibration circuitry minimize the power consumption at a given reliability level, based on experimental models of hardware behavior at each voltage level. The fidelity requirements of each code/data segment can be estimated using feedback optimization tools.

There have been emerging hardware designs that allow operation in less-than-perfect reliability levels. These are discussed in Related Work (Section V). However, to the best of our knowledge, there is a lack of research in understanding how errors in different code/data segments under these reduced reliability conditions impact the overall accuracy of the end result. In this paper we explore the idea of exposing the elastic fidelity requirements of software components to the hardware layer, in order to reduce power and conserve energy while maintaining accuracy guarantees.

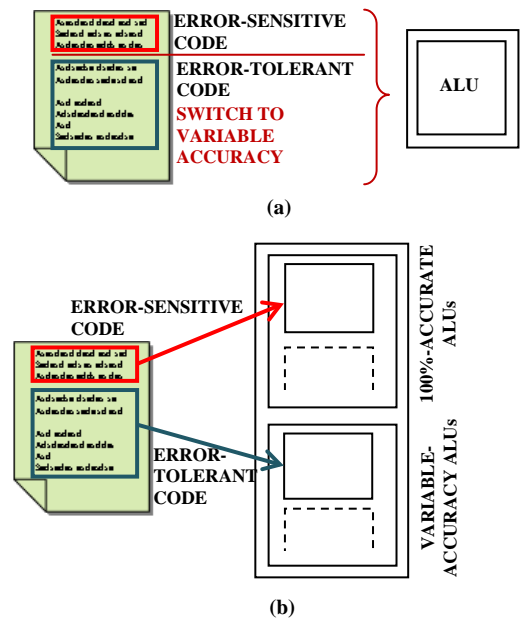

**Figure 1.** Implementing elastic fidelity using (a) a single ALU, and (b) multiple ALUs

## 3. Experimental Methodology

We run multimedia kernels from the MediaBench I and II benchmarking suites [6, 10] on an x86 multicore server, and simulate elastic-fidelity ALUs by injecting errors in the computations at run time through software wrappers. Table 1 lists the applications we use in our experiments. G.721-D performs audio decompression of the G.721 ADPCM speech codec, JPEG-D decompresses a JPEG picture, and H.263-D decompresses compressed video [6]. To judge the quality of the computation, we use segmented signal to noise ratio (SNRseg) for G.721 and peak signal to noise ratio (PSNR) for JPEG and H.263 [27, 35].

Table 1. Application Kernels & Quality Metrics

| Kernel | Application | Quality Metric |
|---|---|---|
| G.721-D | Audio decompression | Segmented Signal to Noise Ratio (SNRseg) |
| JPEG-D | Image decompression | Peak Signal to Noise Ratio (PSNR) |
| H.263-D | Video decompression | Peak Signal to Noise Ratio (PSNR) |

For these applications, we take a raw sample, and run it through the respective encoder to get a compressed file. This compressed file is then decoded using the error-injected kernel to get its decompressed equivalent. This file is compared with the original sample to obtain the quality metric.

### 3.1 Error Injection

We model hardware errors pertaining only to the ALU on a 32-bit data width. This is implemented by injecting errors on the application at run time through software implemented fault injection (SWIFI) using software wrappers [18, 34]. These software wrappers model hardware errors by flipping bits on the 32-bit bus at a given probability (error rate). However, as discussed in Section II, we exclude pointer and branch operations from injecting errors to ensure program stability.

The use of software wrappers allows flexibility in inserting errors in selective locations and helps us to study the behavior of the application not only when errors are injected in the entire computation, but also when they are injected during the execution of specific functions. To gain finer granularity, we vary the bit positions we flip. For example, an application might behave completely differently when errors occur in the 16 most significant bits (MSBs) than when they occur in the 16 least significant bits (LSB) on the ALU. These observations are discussed further in Section IV. For every condition, we run our experiments 1000 times to get statistically significant results.

### 3.2 Metrics for Analysis and Quality Comparisons

To analyze the behavior of the applications, we perform our experiments first by modifying the range of bits to be subjected to error, and then by varying the error probability (error rate).

*1) Bits Subjected to Error:* Starting from the least significant bit, we systematically increase the range of error injection towards the most significant bit, covering the entire 32-bit data width of the ALU. The error probability is kept at 4%. This gives a comprehensive view of the susceptibility of the application to the location of the errors in the bus. In most cases, errors in the least significant bits affect the operation of the program less significantly than those in the most significant bits. In addition, as the scope of error in the bus increases, the error in the computation might go beyond the tolerance of the application, leading it to terminate abruptly. To capture this behavior, we obtain two metrics – number of successful runs, and output quality (SNRseg for the audio, and PSNR for the image and video decoding applications).

*2) Error Rate:* By analyzing the error susceptibility of the application from the prior metrics, we select an optimal range of the bits to allow errors on. This is done to ensure that the program completes successfully at each run. In this case, the error rate is varied to obtain the quality of the output (SNRseg or PSNR). This illustrates the behavior of the application with increasing hardware error rates. The metrics are obtained for errors injected not only in the entire application but also in individual functions to determine their tolerances.

## 4. Results & Analysis

### 4.1 The Impact of Bit Selection on Successful Runs

Fig 2 (a), (d), and (g) plots the number of successful runs (normalized to the error-free execution) for the test kernels. The graphs show the behavior of the entire application as well as when injecting errors only within major functions. In the case of G.721-D, every function runs successfully, even in the presence of errors. However, this is not the case for JPEG-D and H.263-D. For JPEG-D, the upsampling function crashes when the $10^{th}$ least significant bit in the data bus is introduced to error; entropy decoding crashes when errors are injected at the $19^{th}$ least significant bit. Similarly, for H.263-D, the motion compensation block is highly sensitive to errors and crashes even when errors are injected in the $2^{nd}$ least significant bit. The Huffman decoding and reconstruction functions follow soon, crashing at the $12^{th}$ and the $14^{th}$ LSB respectively, while inverse DCT is error-resilient and causes crashes only after the $29^{th}$ bit is flipped.

### 4.2 The Impact of Bit Selection on Output Quality

Fig 2 (b), (e), and (h) plots the decoding quality (SNRseg or PSNR) for the entire application and respective functions. In the case of JPEG-D and H.263-D, we see an anticipated behavior of decrease in the output quality as the higher significant bits are subjected to error. G.721-D shows a peculiar behavior in the fact that the SNRseg decreases up to bit 16 and then starts rising. This anomaly is due to the fact that most of the computations are performed with short int (16 bits) data type. As we increase the scope of error beyond the 16bits, the individual probability of bitflips for the first 16 bits decreases, thus reducing the effect on the computations.

From the results discussed above, we select the range of 8 least significant bits (bit positions 0-7) to be subjected to error for our preliminary analysis of the application behavior with respect to increasing error rates. This is done for two reasons – first errors in the first 8 bits do not affect the output quality significantly and keep it at acceptable levels (> 10dB SNRseg for G.721, > 25dB PSNR for JPEG, and > 30dB PSNR for H.263). Second, the programs run stably without crashing with errors in this region.

Additionally, we observe that entropy decoding in JPEG-D and motion compensation in H.263-D reduce the output quality drastically when injected with errors. Thus, we choose to provide full accuracy to these functions to get a meaningful output. While graphing the behavior of the entire application (curve marked "All"), we exclude error injection in entropy decoding and motion compensation while the remaining application is subjected to the same error rate.

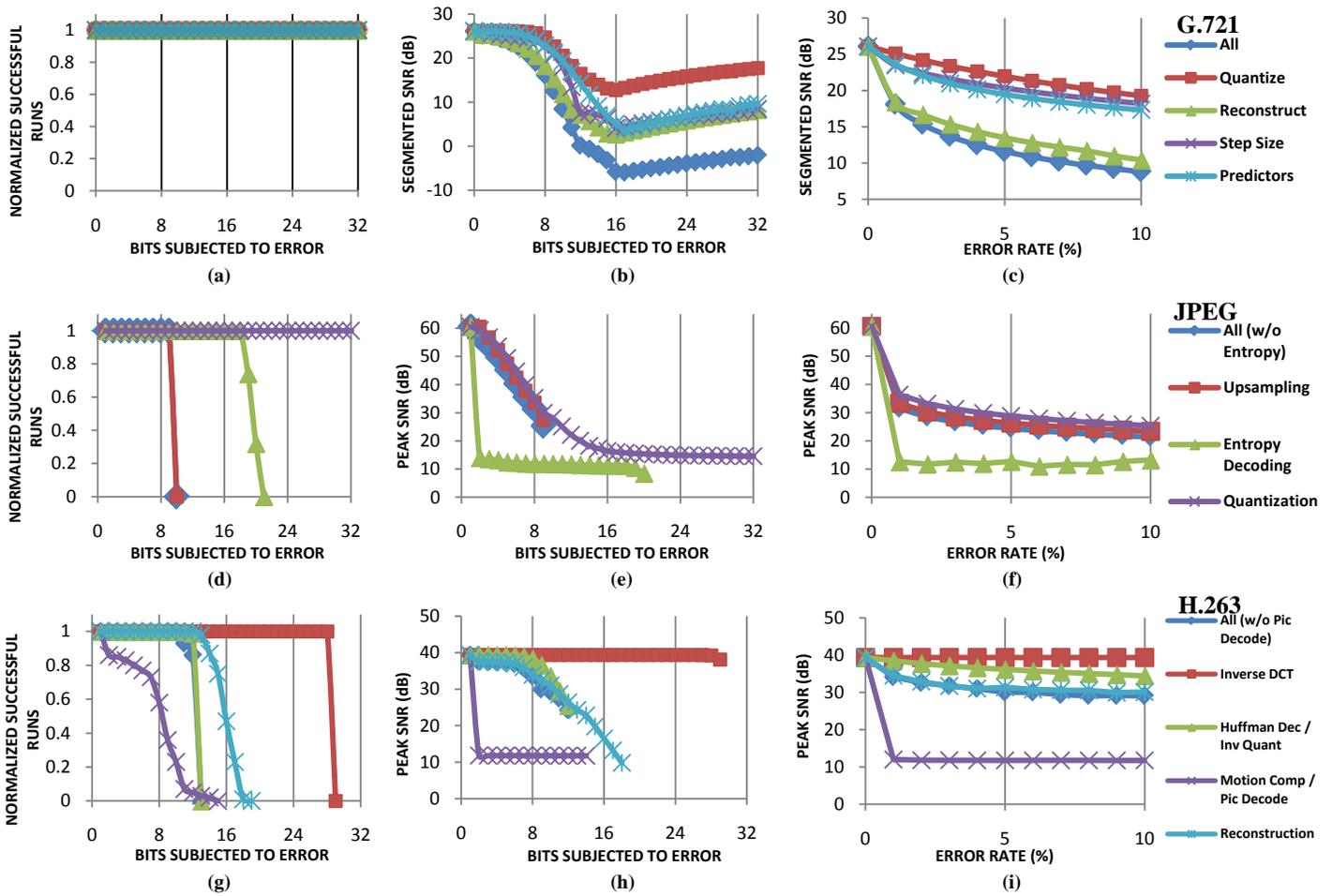

**Figure 2.** (a,d,g) Successful runs vs. bits injected with errors; (b,e,h) Output quality vs. bits injected with errors; (c,f,i) Output quality vs. error rate for G.721 (a,b,c), JPEG (d,e,f) and H.263(g,h,i)

### 4.3 The Impact of Error Rate on Output Quality

Fig 2 (c), (f), and (i) plots the decoding output quality for each application with increasing error rate. In the case of G.721-D, the quantization, step size and predictor functions are more error-tolerant than the reconstruction function. For JPEG-D, upsampling and quantization show similar tolerance with entropy decoding having the lowest. Finally in the case of H.263-D, inverse DCT is not affected by errors in the range of bit positions we consider, while motion compensation is highly sensitive.

### 4.4 Power Savings

As seen from the graphs in Fig 2 (c), (f), and (i), the maximum error rate for acceptable output quality is 7% for G.721-D (SNRreg > 10dB), 4% for JPEG-D (PSNR > 25dB) and 6% for H.263-D (PSNR > 30dB). This is for the case when the entire application is subjected to a uniform error rate. However, by adding variable fidelity levels to the different function routines, Elastic Fidelity Computing can withstand even lower accuracies (higher error rate) for the same output quality level. After determining the corresponding error levels for each major segment of each application, we estimate the power consumption by equating the percentage of dynamic instructions subjected to error [11] with the hardware error-power model presented in [19]. The power consumption is normalized to the power of the processor in error-free operation.

**Table 2.** Estimation of Power Consumption

| Application @ error rate per function | Instructions subjected to error | Normalized processor power consumption |
|---|---|---|
| G.721 audio decoding<br>*Quantization @10%, 0-15 bits;*<br>*Step Size @10%, 0-7 bits;*<br>*Predictors @10%, 0-7 bits;*<br>*all else @4%, 0-7 bits* | 80.7% | 0.89 |
| JPEG image decoding<br>*Entropy @0%;*<br>*Quantization @6%, 0-7 bits;*<br>*Upsampling @5%, 0-7 bits;*<br>*all else @4%; 0-7 bits* | 70.5% | 0.88 |
| H.263 video decoding<br>*Motion compensation @0%;*<br>*IDCT @10%, 0-28 bits;*<br>*all else @6%; 0-7 bits)* | 64.2% | 0.87 |

From Table 2, it is evident that even without any modification to the program, and even when allowing only the ALUs to exhibit variable fidelity, Elastic Fidelity Computing lowers the power consumption of the processor by 11-13%. Because the execution time does not change for these applications when errors are injected in the corresponding functions, the power savings also correspond to energy savings. As our results are highly conservative, we envisage the power savings to be significantly higher once the application is written in a fidelity-aware approach with the necessary program constructs, and when more functional and storage units employ elastic fidelity (rather than only ALUs).

## 5. Related Work

There has been in depth research in modeling the behavior of hardware due to voltage scaling and process variability [7, 9, 28], along with techniques to prevent them [2, 5, 30]. New designs based on better-than-worst-case (BWTC) scenarios have been developed, which relax the design guardbands and perform error recovery when required. Designs such as Razor [8] perform error correction at the hardware through the use of additional circuitry, while [14] corrects the errors at the algorithmic level. The benefits of these techniques in the scope of power reduction are limited due to their error recovery overheads. Moreover, they are orthogonal to Elastic Fidelity and can be used synergistically with it to lower the power and energy consumption even further. In addition, there has been recent work in optimizing the most frequently exercised paths in hardware at the cost of timing errors in the infrequent ones [29]. Techniques like BlueShift [12] work on this principle and optimize the circuit for maximum operating frequency for a given error rate. However, these developments focus on error correction rather than error tolerance and none of them look into the idea of allowing the errors to propagate into the software.

The most significant work related to Elastic Fidelity appears in a related project [19, 20, 21] which targets processor designs that keep voltage-reliability trade-offs in mind. This study minimizes processor power for a given error rate. This differs from our work, as it deals with designing the underlying hardware, while we focus on the feasibility of software to take advantage of this phenomenon.

On the software side, a considerable amount of research has been performed on the effect of single event upsets on software behavior [3, 16, 23, 25, 31] due to the rising reliability issues resulting from decreasing feature size. However, we find that not much has been done in the case of continued errors as presented in this paper. Unlike single event upsets, these errors occur continuously due to faulty hardware (as a result of voltage over-scaling in this case.)

Finally, research on empathic systems [32, 33] considers human perception and user satisfaction to guide power optimizations. Contrary to our work, empathic systems do not trade-off accuracy for power; rather, they trade-off user satisfaction for power. However, similarly to empathic systems, our output quality metrics for the applications we study in this paper are also exploiting human perception to arrive at a result that is good enough, but not necessarily "perfect".

## 6. Conclusion

We observe that not all computations and data in a workload need to maintain 100% fidelity. Our results indicate that some functions in an application are far more error resilient than others. Similarly, errors in certain portions of an application's dataset may cause virtually no change in its operation, while errors in other portions may affect the final outcome significantly (e.g., errors in LSB vs. MSB bits respectively). Elastic Fidelity Computing exploits the variable accuracy requirements within an application to vary the reliability of the underlying hardware according to the application needs at each point in time. Portions of the application that are error-sensitive execute at full reliability, while the ones that are error-tolerant run on variable accuracy to produce an acceptable result to the end user. In turn, the hardware can let go the conservative guardbands imposed on the design to guarantee reliable execution. Instead, hardware may operate at voltage levels low enough to induce errors, but high enough to maintain reasonable output quality.

Our results indicate that Elastic Fidelity Computing can lower the power and energy consumption of workloads pertaining to human perception by 11-13%, while keeping their output quality within acceptable levels. Our results are very conservative, as we assume no modifications to the software, and we assume that only ALUs employ elastic fidelity. We anticipate that by incorporating error-aware design in software, and by extending elastic fidelity operation to more hardware components in a system, the power and energy savings would be significantly higher. Thus, by striking a balance between computational accuracy and supply voltage, and through software/hardware cooperation, Elastic Fidelity Computing shows promise in successfully tackling the ongoing power crisis in processor design.

## References

[1] S. Borkar, "Tackling variability and reliability challenges," *Design & Test of Computers, IEEE,* vol. 23, no. 6, pages. 520-520, 2006.
[2] T. D. Burd and R. W. Brodersen. Design issues for Dynamic Voltage Scaling. In *Proceedings of the 2000 International Symposium on Low Power Electronics and Design.(ISLPED '00),*. pages. 9-14, 2000.
[3] J. Carreira, H. Madeira, and J. G. Silva, "Xception: a technique for the experimental evaluation of dependability in modern computers," *IEEE Transactions on Software Engineering,* vol. 24, no. 2, pages. 125-136, 1998.
[4] K. Chakraborty, P. M. Wells, and G. S. Sohi. Computation spreading: employing hardware migration to specialize CMP cores on-the-fly. In *Proceedings of the 12th international conference on architectural support for programming languages and operating systems (ASPLOS)*, pages. 283-292, 2006.
[5] M. R. Choudhury and K. Mohanram. Masking timing errors on speed-paths in logic circuits. In *Design, Automation & Test in Europe Conference & Exhibition,(DATE '09)*, pages. 87-92, 2009.
[6] L. Chunho, M. Potkonjak, and W. H. Mangione-Smith. MediaBench: a tool for evaluating and synthesizing multimedia and communications systems. In *Proceedings of the Thirtieth Annual IEEE/ACM International Symposium on Microarchitecture (MICRO-30)*, pages. 330-335, 1997.
[7] M. de Kruijf, S. Nomura, and K. Sankaralingam. A unified model for timing speculation: Evaluating the impact of technology scaling, CMOS design style, and fault recovery mechanism. In *Proceedings of the 2010 IEEE/IFIP International Conference on Dependable Systems and Networks (DSN)* , pages. 487-496, 2010.
[8] D. Ernst, K. Nam Sung, S. Das, S. Pant, R. Rao, P. Toan, C. Ziesler, D. Blaauw, T. Austin, K. Flautner, and T. Mudge. Razor: a low-power pipeline based on circuit-level timing speculation. In *Proceedings of the 36th Annual IEEE/ACM International Symposium on Microarchitecture (MICRO-36)*, pages. 7-18, 2003.
[9] F. Firouzi, M. E. Salehi, F. Wang, and S. M. Fakhraie, "An accurate model for soft error rate estimation considering dynamic voltage and frequency scaling effects," *Microelectronics Reliability,* In Press, 2010.
[10] J. E. Fritts, F. W. Steiling, J. A. Tucek, and W. Wolf, "MediaBench II video: Expediting the next generation of video systems research," *Microprocessors and Microsystems,* vol. 33, no. 4, pages. 301-318, 2009.
[11] S. L. Graham, P. B. Kessler, and M. K. Mckusick, "Gprof: A call graph execution profiler," *SIGPLAN Not.,* vol. 17, no. 6, pages. 120-126, 1982.
[12] B. Greskamp, W. Lu, U. R. Karpuzcu, J. J. Cook, J. Torrellas, C. Deming, and C. Zilles. Blueshift: Designing processors for timing speculation from the ground up. In *Proceesings of the 15th IEEE International Symposium on High Performance Computer Architecture (HPCA),* pages. 213-224, 2009.
[13] S. Harizopoulos and A. Ailamaki, "StagedDB: Designing Database Servers for Modern Hardware," *IEEE Data Eng. Bull.,* vol. 28, no. 2, pages. 11-16, 2005.
[14] R. Hegde and N. R. Shanbhag. Energy-efficient signal processing via algorithmic noise-tolerance. In *Proceedings of the* 1999 *International*

*Symposium on Low Power Electronics and Design (ISLPED)*, pages. 30-35, 1999.
[15] M. Horowitz, E. Alon, D. Patil, S. Naffziger, K. Rajesh, and K. Bernstein. Scaling, power, and the future of CMOS. In *Electron Devices Meeting. IEDM Technical Digest. IEEE International*, pages. 7 pp.-15, 2005.
[16] V. Wong, M. Horowitz. Soft Error Resilience of Probabilistic Inference Applications. In *Proceedings of the Workshop on System Effects of Logic Soft Errors (SELSE),* 2006.
[17] C. Hu, "Future CMOS scaling and reliability," *Proceedings of the IEEE,* vol. 81, no. 5, pages. 682-689, 1993.
[18] A. Johansson. Software Implemented Fault Injection Used for Software Evaluation. in *Building Reliable Component-Based Systems*, I. Crnkovic and M. Larsson, Eds., ed: Artech House, 2002.
[19] A. B. Kahng, K. Seokhyeong, R. Kumar, and J. Sartori. Designing a processor from the ground up to allow voltage/reliability tradeoffs. In *Proceedings of the 16th IEEE International Symposium on High Performance Computer Architecture (HPCA)*, pages. 1-11, 2010.
[20] A. B. Kahng, K. Seokhyeong, R. Kumar, and J. Sartori. Recovery-driven design: A power minimization methodology for error-tolerant processor modules. In *Proceedings of the 47th ACM/IEEE Design Automation Conference (DAC)*, pages. 825-830, 2010.
[21] A. B. Kahng, K. Seokhyeong, R. Kumar, and J. Sartori. Slack redistribution for graceful degradation under voltage overscaling. In *Proceedings of the 15th Asia and South Pacific Design Automation Conference (ASP-DAC)*, pages. 825-831, 2010.
[22] J. Kwangok, A. B. Kahng, and K. Samadi, "Impact of Guardband Reduction On Design Outcomes: A Quantitative Approach," *In IEEE Transactions on Semiconductor Manufacturing,* vol. 22, no. 4, pages. 552-565, 2009.
[23] X. Li and D. Yeung. Application-Level Correctness and its Impact on Fault Tolerance. In *Proceedings of the 13th IEEE International Symposium on High Performance Computer Architecture (HPCA)*, pages. 181-192, 2007.
[24] A. Mallik and G. Memik. A Case for Clumsy Packet Processors. In *Proceedings of the 37th annual IEEE/ACM International Symposium on Microarchitecture (MICRO-37)*, pages. 147-156, 2004.
[25] S. S. Mukherjee, C. Weaver, J. Emer, S. K. Reinhardt, and T. Austin. A systematic methodology to compute the architectural vulnerability factors for a high-performance microprocessor. In *Proceedings of the 36th Annual IEEE/ACM International Symposium on Microarchitecture (MICRO-36)*, pages. 29-40, 2003.
[26] J. Patel. CMOS Process Variations: A Critical Operation Point Hypothesis. ed. Computer Systems Colloquium: Stanford University, 2008.
[27] S. R. Quackenbush, T. P. Barnwell, and M. A. Clements. *Objective measures of speech quality*. Englewood Cliffs, N.J.: Prentice Hall, 1988.
[28] D. Roberts, T. Austin, D. Blauww, T. Mudge, and K. Flautner. Error analysis for the support of robust voltage scaling. In *Proceedings of the Sixth International Symposium on Quality of Electronic Design (ISQED)*, pages. 65-70, 2005.
[29] S. Sarangi, B. Greskamp, A. Tiwari, and J. Torrellas. EVAL: Utilizing processors with variation-induced timing errors. In Proceedings of the *41st IEEE/ACM International Symposium on Microarchitecture (MICRO-41)*, pages. 423-434, 2008.
[30] R. A. Shafik, B. M. Al-Hashimi, and K. Chakrabarty. Soft error-aware design optimization of low power and time-constrained embedded systems. In *Design, Automation & Test in Europe Conference & Exhibition (DATE),* pages. 1462-1467, 2010.
[31] P. Shivakumar, M. Kistler, S. W. Keckler, D. Burger, and L. Alvisi. Modeling the effect of technology trends on the soft error rate of combinational logic. In *Proceedings of the 2002 International Conference on Dependable Systems and Networks (DSN)*, pages. 389-398, 2002.
[32] A. Shye, B. Scholbrock, and G. Memik. Into the wild: studying real user activity patterns to guide power optimizations for mobile architectures. In *Proceedings of the 42nd Annual IEEE/ACM International Symposium on Microarchitecture*, pages. 168-178, 2009.
[33] A. Shye, P. Yan, B. Scholbrock, J. S. Miller, G. Memik, P. A. Dinda, and R. P. Dick. Power to the people: Leveraging human physiological traits to control microprocessor frequency. In *Proceedings of the 41st IEEE/ACM International Symposium on Microarchitecture (MICRO-41)*, pages. 188-199, 2008.
[34] R. R. Some, W. S. Kim, G. Khanoyan, L. Callum, A. Agrawal, and J. J. Beahan. A software-implemented fault injection methodology for design and validation of system fault tolerance. In *Proceedings of the 2001 International Conference on Dependable Systems and Networks (DSN 2001)* pages. 501-506, 2001.
[35] T. Veldhuizen. "Measures of image quality," 2010; http://homepages.inf.ed.ac.uk/rbf/CVonline/LOCAL_COPIES/VELDHUIZEN/node18.html.
[36] L. A. Zadeh, "Fuzzy logic, neural networks, and soft computing," *Commun. ACM,* vol. 37, no. 3, pages. 77-84, 1994.
[37] L. A. Zadeh, "Some reflections on soft computing, granular computing and their roles in the conception, design and utilization of information/intelligent systems," *Soft Computing - A Fusion of Foundations, Methodologies and Applications,* vol. 2, no. 1, pages. 23-25, 1998.